# Going From Molecules to Genomic Variations to Scientific Discovery: Intelligent Algorithms and Architectures for Intelligent Genome Analysis

Mohammed Alser[1,*], Joel Lindegger, Can Firtina, Nour Almadhoun, Haiyu Mao, Gagandeep Singh, Juan Gomez-Luna, Onur Mutlu[2,*]

ETH Zurich, Gloriastrasse 35, 8092 Zürich, Switzerland
[1]alserm@ethz.ch, [2]omutlu@gmail.com
*Corresponding authors

**Abstract**

We now need more than ever to make genome analysis more intelligent. We need to read, analyze, and interpret our genomes not only quickly, but also accurately and efficiently enough to scale the analysis to population level. There currently exist major computational bottlenecks and inefficiencies throughout the entire genome analysis pipeline, because state-of-the-art genome sequencing technologies are still not able to read a genome in its entirety. We describe the ongoing journey in significantly improving the performance, accuracy, and efficiency of genome analysis using intelligent algorithms and hardware architectures. We explain state-of-the-art algorithmic methods and hardware-based acceleration approaches for each step of the genome analysis pipeline and provide experimental evaluations. Algorithmic approaches exploit the structure of the genome as well as the structure of the underlying hardware. Hardware-based acceleration approaches exploit specialized microarchitectures or various execution paradigms (e.g., processing inside or near memory) along with algorithmic changes, leading to new hardware/software co-designed systems. We conclude with a foreshadowing of future challenges, benefits, and research directions triggered by the development of both very low cost yet highly error prone new sequencing technologies and specialized hardware chips for genomics. We hope that these efforts and the challenges we discuss provide a foundation for future work in making genome analysis *more intelligent*.
The analysis script and data used in our experimental evaluation are available at: https://github.com/CMU-SAFARI/Molecules2Variations

## 1. Introduction

Sequencing genomic molecules stimulates research and development in biomedicine and other life sciences through its rapidly growing presence in clinical medicine [1–5], outbreak tracing [6–10], and understanding of pathogens and urban microbial communities [11–15]. These developments are driven in part by the successful sequencing of the human genome [16] and in part by the introduction of high-throughput sequencing technologies that have dramatically reduced the cost of DNA sequencing [17]. The bioinformatics community has developed a multitude of software tools to leverage increasingly large and complex sequencing datasets [18–20]. These tools have reshaped the landscape of modern biology and become an essential component of life sciences [21]. The increasing dependence of biomedical scientists on these powerful tools creates a critical need for faster and more efficient computational tools. Our understanding of genomic molecules today is affected by the ability of modern computing technology to quickly and accurately determine an individual's entire genome. In order to computationally analyze an organism's genome, the DNA molecule must first be converted to digital data in the form of a string over an alphabet of four letters or *base-pairs* (bp), commonly denoted by A, C, G, and T. The four letters in the DNA alphabet correspond to four chemical bases, adenine, cytosine, guanine, and thymine, respectively, which make up a DNA molecule. After more than 7 decades of continuous attempts (since 1945 [22]), there is still no sequencing technology that can read a DNA molecule in its entirety. As a workaround, sequencing machines generate randomly sampled subsequences of the original genome sequence, called *reads* [23]. The resulting reads lack information about their order and corresponding locations in the complete genome. Software tools, collectively known as *genome analysis tools*, are used to reassemble read fragments back into an entire genome sequence and infer genomic variations that make an individual genetically different from another.

There are five key initial steps in a standard genome sequencing and analysis workflow [18], as we show in **Figure 1**. The first step is obtaining the genomic data either through sequencing a DNA molecule, downloading real data from publicly available databases, or computer simulation. Sequencing requires the collection and preparation of the sample in the laboratory or on-site. The second step, known as *basecalling*, is to process the raw sequencing data as each sequencing technology generates different representations of the sequencing data. The basecalling step must convert the raw sequencing data into standard format of sequences of A, C, G, and T in the DNA alphabet. The third step, known as *quality control*, examines the quality of each sequenced base and decides which bases of a read to trim, as sequencing machines introduce different types and rates of sequencing errors leading to inaccurate end results. The fourth step is a process known as *read mapping*, which matches each read sequence to one or more possible locations within the reference genome (i.e., a representative genome sequence for a particular species), based on the similarity between the read and the reference sequence segment at that location. Unfortunately, the bases in a read may not be identical to the bases in the reference genome at the location that the read actually comes from. These differences may be due to (1) sequencing errors (with a rate in the range of 0.1-20% of the read length, depending on the sequencing technology), and (2) genetic differences that are specific to the individual organism's DNA and may not exist in the reference genome. A read mapping step must tolerate such differences during similarity check, which makes read mapping even more challenging. The fifth step, known as *variant calling*, aims to identify the possible genetic differences between the reference genome and the sequenced genome. Genetic differences include small variations [24] that are less than 50 bp, such as single nucleotide polymorphisms (SNPs) and small insertions and deletion (indels). Genetic differences can also be larger than 50 bp variations, known as structural variations [25], which are caused by chromosome-scale changes in a genome. For example, insertion of an about 600,000-base long region has been observed in some chromosomes [26].

## 1 Obtaining Genomic Sequencing Data

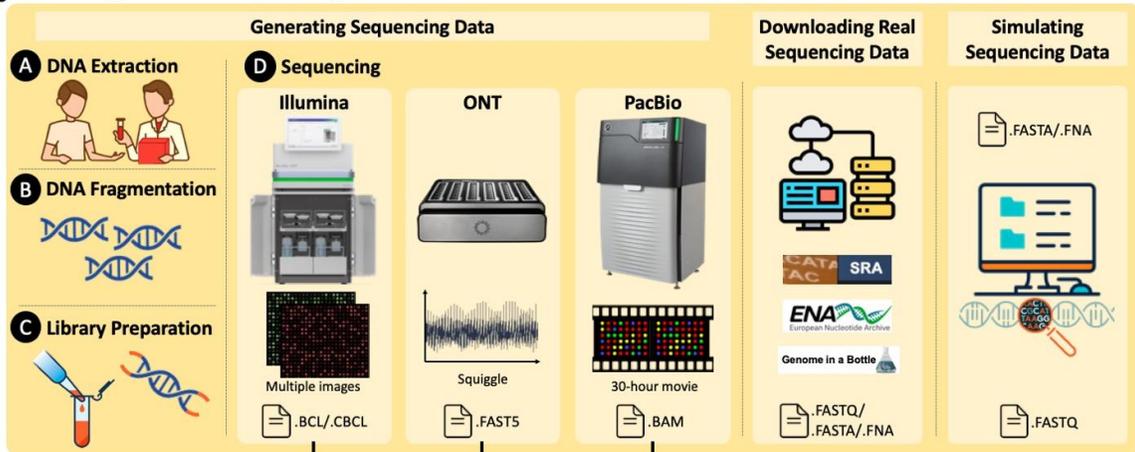

## 2 Basecalling

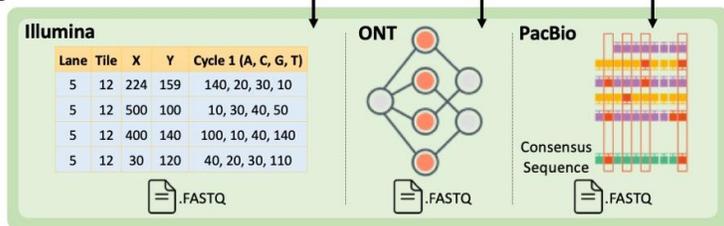

## 3 Quality Control

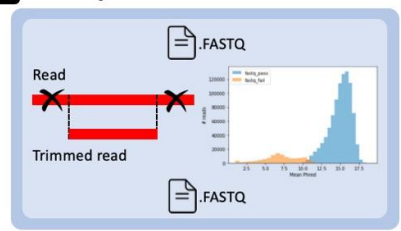

## 4 Read Mapping

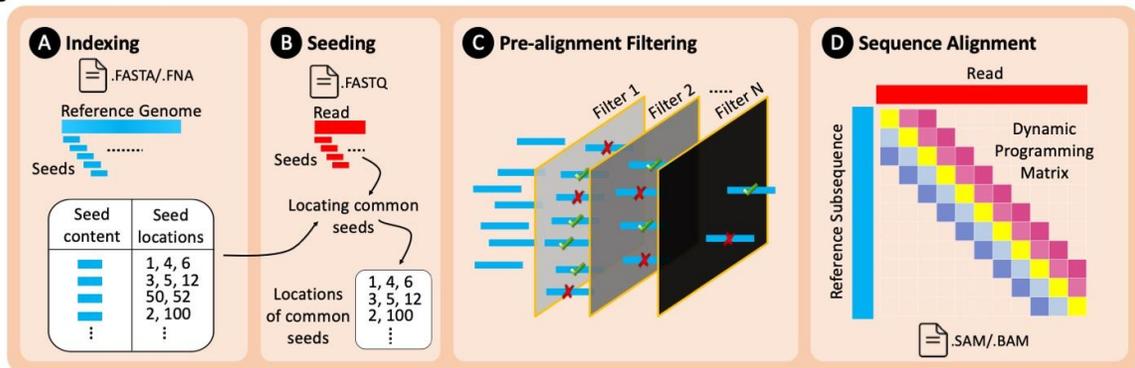

## 5 Variant Calling

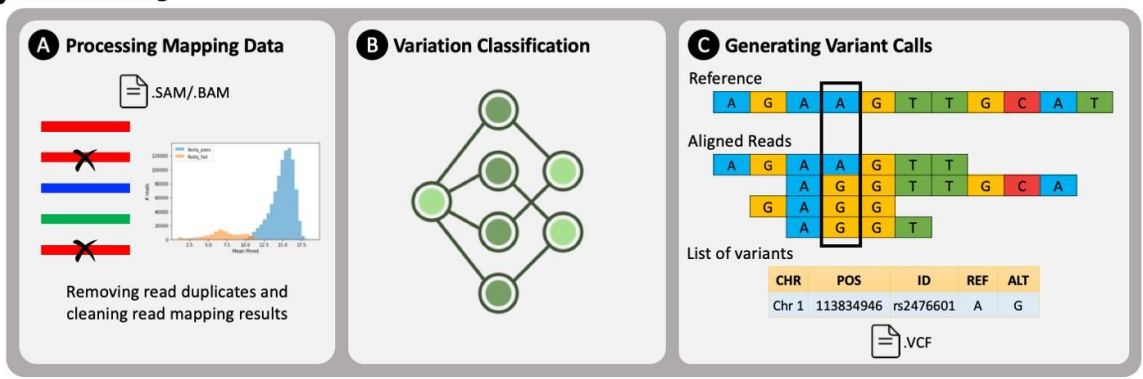

**Figure 1: Overview of a typical genome analysis pipeline. 1)** Genomic sequencing data is first obtained through sequencing a new sample, downloading from publicly-available databases, or computer simulation. Sequencing starts with (A) extracting the DNA, (B) fragmenting it, (C) preparing its library, and (D) using sequencing machines for providing raw sequencing data. **2)** Raw sequencing data needs to be converted into read sequences of A, C, G, T in the DNA alphabet using basecalling techniques. Basecalling techniques are sequencing technology dependent. **3)** To ensure high quality sequencing reads, a quality control step is performed to filter out low quality subsequences of a read or an entire read sequence. **4)** Read mapping step is performed to locate each read sequence in a reference genome. Read mapping is four steps: (A) indexing the reference genome, (B) extracting seeds from each read and locating common seeds with the index, (C) pre-alignment filtering dissimilar sequence pairs, and (D) performing sequence alignment for every sequence pair that passes the filtering. **5)** Detecting and inferring genomic variations usually consists of three steps: (A) processing the read mapping data for increasing its quality, (B) classifying variations between mapped reads and a reference genome, and (C) identifying genomic variations.

We define **intelligent genome analysis** as the ability to read, analyze, and interpret genomes not only quickly, but also accurately and efficiently enough to scale the analysis to population level. Some existing works on accelerating one or more steps in genome analysis sacrifice the optimality of the analysis results in order to reduce execution time (as described in [18,27]). Genome analysis is currently a first-tier diagnostic test for critically ill patients with rare genetic disorders, which necessitates the need for making the analysis much faster while maintaining the same or providing better analysis accuracy for successful clinical practice [1,2,28,29]. For example, it is observed that at least 10% of read sequences simulated from the human reference genome remain unmapped across 14 state-of-the-art aligners [30] due to potential mapping artifacts [31], which demonstrates that accuracy is still an issue even in read mapping that is extensively studied. On the other hand, the vast majority of genome analysis tools are implemented as software running on general-purpose computers, while sequencing is performed using extremely specialized, high-throughput machines [18]. This introduces two main problems. 1) Modern sequencing machines generate read fragments at an exponentially higher rate than prior sequencing technologies, with their growth far outpacing the growth in computational power in recent years [32]. 2) Genome analysis generates a large amount of *data movement* between the sequencing machine and the computer performing the analysis and between different components (e.g., compute units and main memory) of computers. The data movement across power-hungry buses is extremely costly in terms of both execution time and energy [33–41]. Increasing the number of CPUs used for genome analysis may decrease the overall analysis time, but significantly increases energy consumption and hardware costs, and potentially worsens the data movement bottleneck (more cores competing for memory access) [42,43]. Cloud computing platforms still suffer from similar issues along with additional concerns due to genomic data protection guidelines in many countries [44–49].

These costs and challenges are a significant barrier to enabling intelligent genome analysis that can keep up with sequencing technologies. As a result, there is a dire need for new computational techniques that can quickly process and analyze a tremendous number of extracted reads in order to drive cutting-edge advances in applications of genome analysis [27,35,36,50,51]. There exists a large body of work trying to tackle this problem by using intelligent algorithms, intelligent hardware accelerators, and intelligent hardware/software co-design [33–35,52–54]. Computer algorithms and hardware architectures are called *intelligent* if they are able to efficiently satisfy three principles, data-centric, data-driven, and architecture/algorithm/data-aware [55]. First, we would like to process genomic data efficiently by minimizing data movement and maximizing the efficiency with which data is handled, i.e., stored, accessed, and processed. Second, we would like to take advantage of the vast amounts of genomic data and metadata to continuously improve decision making (self-optimizing decisions) for many different use cases in science, medicine, and technology. Third, we would like to orchestrate the multiple components across the entire analysis system and adapt algorithms by understanding the structure of the underlying hardware, understanding analysis algorithms, and understanding various properties (i.e., the structure of the genome, type of sequencing data, quality of sequencing data) of each piece of data. Our **goal** in this work is to survey a prominent set of these three types of intelligent acceleration efforts for guiding the design of new highly-efficient tools for intelligent genome analysis. *To this end*, we (1) discuss various state-of-the-art mechanisms and techniques that improve the execution time of one or more steps of the genome analysis pipeline using different modern high-performance computing architectures, and (2) highlight the challenges, that system designers and programmers must address to enable the widespread adoption of hardware-accelerated intelligent genome analysis.

## 2. Obtaining Genomic Sequencing Data

Genomic sequencing data (reads) can be obtained by 1) sequencing a DNA sample, 2) downloading real sequencing data from publicly available databases, or 3) simulating sequencing data, as we show in **Figure 1.1**.

### 2.1. Generating Sequencing Data

One of the earliest successful protein sequencing attempts was made in the 1950s by Frederick Sanger who devoted his scientific life to the determination of the chemical structure of biological molecules, especially that of insulin [22,23,56,57]. The first DNA sequencing was successful only after two decades, in 1977, as introduced in two different sequencing methods by Sanger and Coulson [56], and by Maxam and Gilbert [58]. Though there was constant progress in sequencing attempts for different small genomes, we could obtain the first draft of the human genome sequence only in June 2000 as a result of a large international consortium, costing more than USD 3 billion (USD 1 for each DNA base) and more than 10 years of research [16,59]. This sequencing era was referred to as *first generation sequencing*. Since then, DNA sequencing has evolved at a fast pace with increasing sequencing throughput and decreasing cost, which lead to frequent updates and major improvements to the human genome sequence [60–62] and very recently have resulted in a near-complete human genome sequence [63]. These advances reshaped the landscape of modern biology and sequencing became an essential component of biomedical research.

Generating sequencing data includes three key steps: sample collection, preparation (known as *library preparation* [64]), and sequencing (**Figure 1.1**). Sample collection and library preparation significantly depend on the protocol, preparation kit, minimum amount of DNA in the sample, and the sequencing machine, which require meeting rigorous requirements by the manufacturer of sequencing machines for the successful sequencing. The sample collection and library preparation steps are performed using non-computational methods in the "wet" laboratory or on-site, especially when the used sequencing machine is portable, prior to performing the actual sequencing. Each sequencing machine has different properties such as sequencing throughput, read length, sequencing error rate, type of raw sequencing data, sequencing machine size, and cost. Sequencing throughput is defined as the number of bases generated by the sequencing machine per second. Reads can have different lengths and they can be categorized into three types: 1) short reads (up to a few hundred bp), 2) ultra-long reads (ranging from hundreds to millions of bp), and 3) accurate long reads (up to a few thousands of bp). Though most of the existing sequencing machines are fundamentally different, they also share common properties, such as requiring library preparation, generating only fragments (i.e., reads) of the DNA sequence, introducing errors in the output sequencing data, and requiring converting the raw sequencing data into sequences of nucleotides (i.e., A, C, G, and T in the DNA alphabet). Most existing sequencing technologies start with DNA fragmentation as part of the library preparation protocol. DNA fragmentation refers to intentionally breaking DNA strands into fragments using, for example, resonance vibration [64,65]. DNA fragmentation helps to exploit a large number of DNA fragments for higher sequencing yield, as sequencing quality usually degrades towards the end of long DNA fragments [66] due to for example the limited lifetime of polymerase enzymes used for sequencing [67].

## 2.2. Downloading Real Sequencing Data

Research groups, laboratories, and authors of research papers normally release their sequencing data on publicly available repositories to meet requirements of both reproducibility in biomedical and life science research and journals for data sharing [68]. There are currently more than 29 peta ($10^{15}$) bases of sequencing data publicly available as FASTQ files on the Sequence Read Archive (SRA) database [69], which is doubling in the number of bases every 2 years [70]. Other databases are also available, such as the European Nucleotide Archive (ENA) [71]. There are also a large number of reference genome sequences, as FASTA files, for more than 108,257 distinct organisms publicly available in the NCBI Reference Sequence Database (RefSeq) database [72], which is doubling in the number of organisms every 3 years [73].

## 2.3. Simulating Sequencing Data

Sequencing data can be generated using computer simulation [74–77]. Many computational laboratories use simulated sequencing data, as they lack adequate resources to generate their sequencing data using sequencing machines [19] or lack access to gold standard experimental data when self-assessing a newly developed tool [12,78]. Existing sequencing technologies use different mechanisms and chemistries that result in sequencing data with different characteristics. Read simulators take into account many of these characteristics by modeling them according to each technology. These simulators differ in target sequencing technology, input requirements, and output format. They also have several aspects in common, such as requiring a reference genome, the minimum and maximum read lengths, sequencing error distribution, and type of genetic variations (e.g., substitutions, insertions, or/and deletions) that make read sequences different from the reference genome sequence. Most read simulators generate different standard file formats, such as FASTQ, FASTA, or BAM, which can be used to locate potential mapping locations of each read in the reference genome.

Examples of read simulators for these three types include ART [79], Mason [77], and ReSeq [80] for short reads, PBSIM2 [75] and NanoSim [76] for ultra-long reads, and PBSIM [81] for accurate long reads (known as HiFi reads). However, the use of read simulations poses several limitations, as simulated data recapitulates the important features of real data and oversimplifies/biases the challenge for tested methods [19]. To avoid such biases, a common practice and more-comprehensive approach is to complement the simulated data with real experimental data. Hence, most journals require making the new sequencing data publicly accessible and explicitly reporting the *accession number* (a unique identifier given to sequencing data) of existing sequencing data used in the study.

## 3. Types of Genomic Sequencing Data

There currently exist three different sequencing technologies that are widely-used to sequence DNA samples around the globe and in space [82]. Current prominent sequencing technologies and their output data can be categorized into three types: 1) short reads, 2) ultra-long reads, and 3) accurate long reads. In **Table 1**, we summarize the main differences between these three types of prominent sequencing technologies. We provide more details on the sequencing machine size, cost, throughput, maximum library preparation time, and maximum sequencing time for the most capable instrument for each sequencing technology.

**Table 1. Summary of the main differences between the three types of prominent sequencing technologies. We choose the most capable instrument as a representative of each sequencing technology.**

|  | **Short Reads** | **Ultra-long Reads** | **Accurate Long Reads** |
|---|---|---|---|
| **Leading company** | Illumina (https://www.illumina.com) | Oxford Nanopore Technologies (https://nanoporetech.com) | Pacific Biosciences (https://www.pacb.com) |
| **Representative instrument** | Illumina NovaSeq 6000 | ONT PromethION 48 | PacBio Sequel IIe |

| Instrument picture | 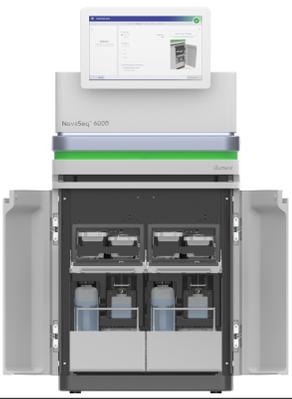 | 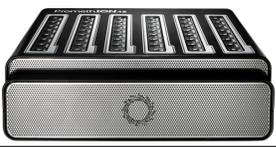 | 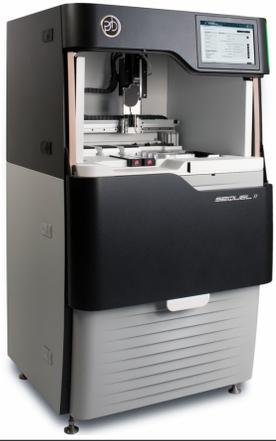 |
|---|---|---|---|
| **Instrument weight** | 481 kg | 28 kg | 362 kg |
| **Instrument dimension** (W×H×D in cm) | 80 × 94.5 × 165.6 | 59 × 19 × 43 | 92.7 x 167.6 x 86.4 |
| **Instrument cost**[1] | 3Y | Y | 1.6Y |
| **Read length** | 100-300 | 100-2M | 10K-30K |
| **Read length in a single data file** | Fixed | Variable | Modest Variability |
| **Accuracy** | 99.9% | 90%-98% | 99.9%[2] |
| **Maximum sequencing throughput per run** | 6 Tb | 14 Tb | 35 Gb (160 Gb of raw sequencing data)[3] |
| **Number of flow cells operating simultaneously** | 2 | 48 | 1 |
| **Hands-on library preparation time** | 45 minutes - 2 hours | 10 minutes-3 hours | 6 hours |
| **Turnaround library preparation time**[4] | 1.5-6.5 hours | 24 hours | 24 hours |
| **Sequencing run time** (including basecalling) | 44 hours | 72 hours | 30 hours |

[1]Y can be as low as USD 300,000. The cost does not include consumables (e.g., flow cells and reagents) needed for each sequencing run.
[2]Consensus accuracy.

[3]Per SMRT cell. Sequel IIe can process up to 8 SMRT cells sequentially, slide 25 https://www.pacb.com/wp-content/uploads/HiFi_Sequencing_and_Software_v10.1_Release_Technical_Overview_for_Sequel_II_System_and_Sequel_IIe_System_Users-Customer-Training-01.pdf.
[4]Includes both hands-on library preparation time and waiting time.

### 3.1. Short Reads

Short read sequencing (sometimes called *second generation sequencing*) technologies, such as Illumina [17,83,84] and Singular Genomics [85], generate subsequences of length 100-300 bp. Illumina is currently the dominant supplier of sequencing instruments. The key advantage of Illumina short reads is the significantly low sequencing error rate (as low as 0.1% of the read length) introduced by the sequencing machine [86]. Over 6 terabases can be generated in a single sequencing run in about two days using a single instrument (Illumina NovaSeq 6000). Illumina sequencing technology, called *sequencing by synthesis* (SBS), is very similar to that of the first generation sequencing (i.e., Sanger) [87,88]. A key procedural difference in comparison to Sanger sequencing is in the preparation of the sequencing library and the degree of parallelism and throughput during sequencing. Sanger sequencing libraries require multiple steps that require growth in culture and DNA isolation before sequencing [88]. This multistep process can be completed in approximately one week, at which point the processed DNAs are ready for sequencing.

Illumina sequencing requires a hands-on time of less than two hours (or a turnaround time of up to 6.5 hours) to prepare the sample for sequencing. Library preparation in Illumina sequencing starts with fragmenting the DNA into short fragments and adding *adapters* (short single strands of synthetic DNA, called oligonucleotides [89]) to both fragment ends. These adapters enable binding each fragment to the flow cell. Fragments can then be amplified to create clusters of up to 1,000 identical copies of each single fragment. Sequencing is then performed base-by-base using a recent technique called *2-channel sequencing* [90]. During each sequencing cycle (3.5-6.75 minutes [91]), a mixture of two nucleotides, adenine and thymine, labeled with the same fluorescent dye (i.e., green color) is added to each cluster in the flow cell. Images are taken of the light emitted from each DNA cluster using a CMOS sensor. Next, a mixture of two nucleotides, adenine and cytosine, labeled with another fluorescent dye (i.e., red color) is added to each cluster in the flow cell. Another image is taken of the light emitted from each DNA cluster. During basecalling, the combinations of "light observed" and "no light observed" in the two images are interpreted. E.g., if the light is only observed in the first image, it is interpreted as a thymine base. If the light is observed in the second image, it is interpreted as a cytosine base. Clusters with light in both images are flagged as adenine bases, while clusters with no light in both images represent guanine bases. This process is repeated on each nucleotide for the length of the DNA fragment (read). The large number of clusters and the straightforward basecalling process make Illumina sequencing provide the highest throughput of accurate bases compared to other sequencing techniques.

### 3.2. Ultra-long Reads

Ultra-long read (or *nanopore*) sequencing, called *third generation sequencing*, is more recent than Illumina sequencing. The first nanopore sequencing machine, MinION, was introduced in 2014 and made commercially available in 2015. The main concept behind nanopore sequencing was brainstormed much earlier in the 1990s [92]. The first MinION sequencing machine was promising as it was incredibly small in size (smaller than the palm of a hand and weighing only 87 g) compared to existing sequencing machines. However, its sequencing accuracy rate was 65% (i.e., one out of every three bases is erroneous, which can significantly degrade the accuracy of downstream analyses) [93].

The nanopore sequencing technology requires first preparing the sequencing library by fragmenting the DNA sequence and adding a sequencing adapter and a motor protein at each end of the fragment. The sequencing starts with passing each DNA fragment through a nanoscale protein pore (nanopore in short) that has an electrical current passing through it. The sequencer measures changes to an electrical current as nucleic acids, each with different electrical resistance, are passed through the nanopore. Using the computational basecalling step, the electrical signals are decoded into a specific DNA sequence. The motor protein helps control the translocation speed of the DNA fragment through the nanopore. Over 14 terabases can be generated in a single sequencing run in about 3 days using a single ONT PromethION 48 instrument (**Table 1**). Recent nanopore machines use dual electrical current sensors (called reader heads) to increase the accuracy of sensing and improve the detection resolution [93,94]. The accuracy of nanopore sequencing technology has been constantly improving from 65% to above 90% and can reach 98% for some reads due to both dual sensing and improved basecalling [95–98]. However, this comes at the cost of computationally-expensive basecalling, as we discuss in **Section 5**.

### 3.3. Accurate Long Reads

The latest sequencing technology, referred to as third or fourth sequencing technology [99], is from Pacific Biosciences (PacBio). It generates high-fidelity (HiFi) reads that are relatively long (10-30K) and highly accurate (99.9%) [95,100,101]. The PacBio sequencing requires fragmenting the DNA molecule and adding double-stranded adapters (called SMRTbell) to both ends of the fragment. The DNA fragment has a DNA subsequence binding to its reverse complement sequence. This creates a circular DNA fragment for sequencing. The PacBio sequencing leverages multiple pass circular consensus sequencing (CCS) by sequencing the same circular DNA fragment at least 30 times and then correcting errors by calculating a consensus sequence [101]. Each sequencing pass through the circular DNA fragment produces a *subread*, which is used to calculate the consensus sequence by overlapping all resulting subreads of a single DNA fragment. The PacBio sequencing uses a polymerase that passes through the circular DNA fragment and incorporates fluorescently labeled nucleotides. As a base is held by the polymerase, fluorescent light is produced and recorded by a camera in real-time. The camera provides a movie of up to 30 hours of continuous fluorescent light pulses that can be interpreted into bases. The PacBio sequencing provides the least sequencing throughput compared to Illumina and ONT. Over 35 gigabases can be generated in a single sequencing run in about 30 hours (**Table 1**).

## 3.4. Discussion on Types of Sequencing Reads

Short reads have the advantages of both low sequencing error rate and high sequencing throughput (number of basecalled bases). Another property of short reads that can be considered as an advantage is the equivalent length of all reads stored in the same FASTQ file, which helps in achieving, for example, load balancing between several CPU threads. Repetitive regions in genomes pose challenges for constructing assembly (de novo assembly [102]) using short reads. De novo assembly is an alternative to read mapping, in which it constructs the sequence of a genome from overlapping read sequences without comparison to a reference-genome sequence. Both ultra-long reads and accurate long reads in general offer better opportunities for genome assembly and detecting complex structural variant calling.

Ultra-long reads provide two main advantages: 1) providing more contiguous assembly than that of short reads, where each contig can read up to 3 Mb (covering the typical length of bacteria genomes) [103], and 2) its read length is theoretically limited only by the length of the DNA fragment translocating through the pore [104] as it does not require enzyme-based nucleotide incorporation, amplification for cluster generation, nor detection of fluorescence signals [93,105]. However, nanopore sequencing data suffers from high sequencing error rate and some of its computational analysis steps require longer executing time and higher memory footprint compared to both short reads and accurate long reads. For example, performing de novo assembly (using Canu [106]) using ultra-long reads is at least fourfold slower than that when using accurate long reads, which is mainly because of the errors in the raw sequencing data [103]. This also leads to introducing new computational steps with the goal of polishing errors in the assembly [96,98,107]. Such new steps normally increase the computation overhead of analyzing ultra-long reads as the new steps are computationally expensive [107]. Another example of expensive steps for analyzing ultra-long reads is basecalling, which is based on neural networks. For this, ONT PromethION 48 includes 4 A100 GPU boards for accelerating basecalling and coping with its sequencing throughput.

The high accuracy of accurate long reads is a key enabler of the recent improvements in human genome assembly and unlocking complex regions of repetitive DNA [63]. More than 50% of the regions previously inaccessible with Illumina short reads for the GRCh37 human reference genome are now accessible with HiFi reads (supplemented with ultra-along reads) for the GRCh38 human reference genome [101]. For other applications, such as profiling microbiomes through metagenomics analysis, the use of accurate long reads leads to detecting the same numbers of species as with the use of short reads [108]. However, the sequencing cost and computational expensive basecalling required to generate HiFi reads currently challenge widespread adoption.

This suggests that there is no preferable sequencing data type for all applications and use cases. Each sequencing technology has its own unique advantages and disadvantages. The short reads will continue to be widely used due to their very high accuracy and low cost. With increases in read lengths of Illumina sequencing technology [109], there will be a growing demand for adjusting existing algorithms or introducing new algorithms that leverage new properties of anticipated Illumina sequencing data. With increases in accuracy of long and ultra-long reads, there will be a growing demand for improving execution time of their computational steps (e.g., basecalling) and reducing overall sequencing cost to enable widespread adoption.

## 4. Genome Analysis Using Different Types of Sequencing Reads

We evaluate the performance of three typical genome analysis pipelines (**Figure 1**) for the three prominent sequencing data types, 1) short reads, 2) ultra-long reads, and 3) accurate long reads, as we show in **Figure 2**. We report the throughput of each step using a single CPU thread, running on a 2.3 GHz Intel Xeon Gold 5118 CPU with up to 48 threads and 192 GB DDR4 RAM. Note that nanopore basecalling is based on the throughput of Guppy running on a CPU [110]. We report the sequencing throughput using a single flow cell. We calculate the throughput of each step by dividing the number of bases (outputted by sequencing and basecalling or taken by quality control, read mapping, and variant calling) over the total execution time in hours. We provide the data and exact command lines used to run each tool in the GitHub repository of this paper.

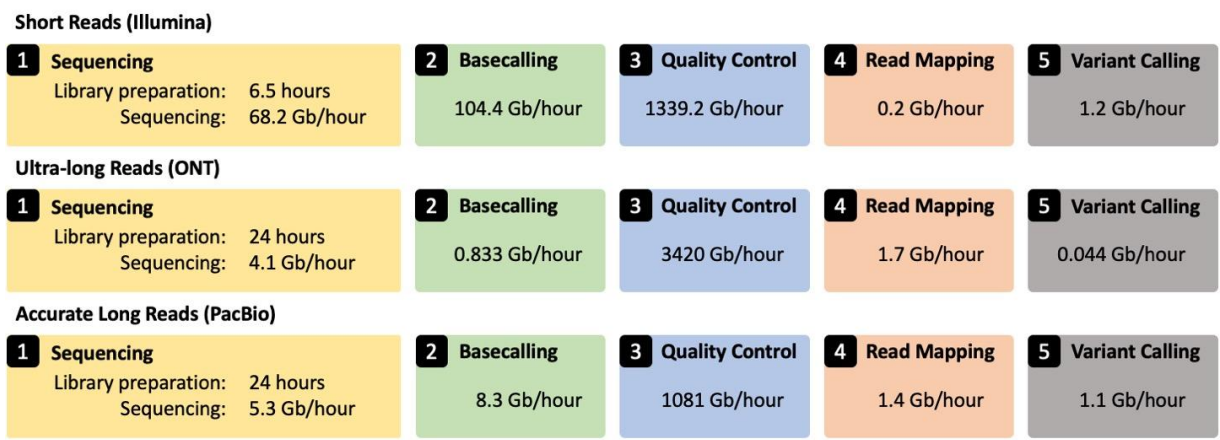

**Figure 2: Performance comparison of the five main steps of genome analysis pipeline.**

We make five key observations based on **Figure 2**. 1) Short read sequencing provides the highest throughput per flow cell compared to other long read sequencing technologies. 2) Genome analysis of both ultra-long reads and accurate long reads suffers from long execution time of their basecalling step. This is expected because both technologies use computationally expensive basecalling steps, as we explain in detail in **Section 5**. 3) Read mapping is the most computationally expensive step, followed by variant calling, in the genome analysis pipeline for short reads. 4) The first major bottleneck in the genome analysis pipelines for both accurate long reads and ultra-long reads is variant calling. Read mapping is the second and the third bottleneck in the genome analysis pipelines for accurate long reads and ultra-long reads, respectively. 5) Sequencing throughput is 341× and 56.8× higher than the throughput of read mapping and variant calling of short reads, respectively. Sequencing throughput is 2.4× and 93.2× (3.8× and 4.8×) higher than the throughput of read mapping and variant calling of ultra-long reads (accurate long reads), respectively.

We conclude that each type of sequencing data imposes different acceleration challenges and creates its own computational bottlenecks. There is also a dire need for developing new computational techniques that can overcome the existing computational bottlenecks and building new hardware architectures that can reduce data movements between different steps of genome analysis and improve overall analysis time and energy efficiency. In the next sections, we survey various state-of-the-art mechanisms and techniques that improve the execution time of one or more steps of the genome analysis pipelines for different types of sequencing data.

## 5. Basecalling

Most existing sequencing machines do not provide read sequences in the DNA alphabets. Instead, they provide a native format that is sequencing technology dependent. Thus, existing sequencing technologies require a *basecalling* step (**Figure 1.2**) to convert the native sequencing data format into a standard format that can be understood by all existing genome analysis tools, regardless of the sequencing technology used. Basecalling is the first computational step in the genome sequencing pipeline that converts raw sequencing data (images for Illumina, movie for PacBio, or electric current for ONT) into sequences of nucleotides (i.e., A, C, G, and T in the DNA alphabet). We provide in this section a brief description of basecalling for the three prominent sequencing technologies, Illumina short reads, Oxford Nanopore ultra-long reads, and PacBio HiFi accurate long reads. We summarize the main differences between their basecalling techniques in **Table 2**.

### 5.1. Illumina

During each cycle of Illumina sequencing, two images are taken to identify the possible chemical reactions occurring in DNA clusters. The light intensities captured in one or both images directly represent the type of the subject nucleotide, as we explain in **Section 3.1**. This information is stored as binary data in the *CBCL* file format. Each sequencing run provides a large number of CBCL files that need to be converted into a single FASTQ file. The basecalling per CBCL file can take up to 0.21 minutes (50% of it is spent on reading and writing from/to files) [111], which is significantly less than the time for a single sequencing cycle (3.5-6.75 minutes [91]).

**Table 2. Summary of the main differences between the three types of prominent sequencing technologies. We choose the most capable instrument as a representative of each sequencing technology.**

|  | **Short Reads** | **Ultra-long Reads** | **Accurate Long Reads** |
|---|---|---|---|
| **Type of raw sequencing data** (before basecalling) | Multiple images of fluorescence intensities for each sequencing cycle | Electrical signal for each DNA segment | Fluorescence traces captured continuously into a 30-hour movie |
| **Input file format for basecalling** | BCL or CBCL | FAST5 | BAM |
| **Expected size of basecalling input file** | One CBCL file of size 350MB per cycle, lane, and surface | 10× the size of the corresponding FASTQ file | Subreads.BAM of size 0.5-1.5 TB |
| **Basecalling algorithm** | BCL2FASTQ | Guppy/Bonito (deep neural networks) | CCS |
| **Basecalling time** | 48 minutes[1] | 142 minutes[2] | 24 hours[3] |
| **Number of basecalled bases** | 83.5 Gb[1] | 20 Gb[2] | 200 Gb of HiFi yield[3] |

[1]BCL2FASTQ based on SRR2890933, 1.67 billion reads (full 8 lanes) at 50bp read length [112].
[2]Using a single V100 GPU, adapted from [110].
[3]Using CCS v6.0 for a 25 KBases library on 2x64 cores at 2.4 GHz, adapted from [113].

### 5.2. ONT

For nanopore sequencing, the conversion of the raw electrical current signal or *squiggle* into a sequence of nucleotides is challenging because: (1) signals have stochastic behavior and low signal-to-noise ratio (SNR) due to thermal noise and the lack of statistically significant current signals triggered by DNA motions, (2) electrical signals can have long dependencies of event data on neighboring nucleotides, and (3) the sensors (reader heads) used cannot measure the changes in electrical current due to a single nucleotide, but rather measuring the effect of multiple nucleotides together. Nanopore raw data are current intensity values measured at 4 kHz, saved in the *FAST5* format (a modified *HDF5* format). Nanopore basecalling is computationally expensive and its algorithm is quickly evolving. Neural networks have supplanted hidden Markov model (HMM) based basecallers for their better accuracy, and various neural network structures are being tested [95,114]. Another important advantage of nanopore basecalling is its ability to improve read accuracy (by 10% [115]) by correcting possible sequencing errors.

ONT provides a production basecaller, called *Guppy*, and its development version, *Bonito*. Guppy is optimized for performance and accelerated using modern GPUs. Modern neural network basecallers such as Guppy and Bonito are typically composed of: (1) residual network with convolutional layers, and/or (2) long short-term memory-based recurrent layers. RNN-based models are used to model the temporal dependencies present in basecalling data. Other than Guppy, which is a GPU-based basecaller, most of the existing basecallers are only implemented for CPU execution. Thus, basecallers lack dedicated hardware acceleration implementations, which could greatly reduce the basecalling time. Even Guppy takes 25 hours to basecall a 3 Gb human genome on a powerful server-grade GPU [116]. Potential alternatives to further accelerate Guppy and other basecallers are processing-in-memory techniques [33–36,50,52–54,117–119], which consist of placing compute capabilities near or inside memory. PIM techniques are particularly well suited for workloads with memory-bound behavior, as for example RNNs show [39,120]. For that, algorithm/architecture co-designed processing-in-memory (PIM) can accelerate Guppy by 6× [116]. ONT also provides Guppy-lite, a throughput-optimized version of Guppy that provides higher performance at the expense of the basecalling accuracy. Therefore, basecallers need to balance the tradeoff between speed and accuracy depending on the requirements of a particular application [121].

Instead of bascalling followed by analysis in basepair space, several works propose analyzing reads directly in their raw signal space, obviating or alleviating the need for expensive basecalling [122,123]. One motivation for such proposals is the *Read-Until* feature [124] of ONT sequencing machines, which allows to physically eject reads from each nanopore in real time if they are deemed not interesting for the application (e.g., do not belong to target species). Ejecting a DNA segment requires analyzing the partial squiggle signal much earlier than completing sequencing the complete DNA segment, which the computationally expensive basecalling cannot satisfy. SquiggleNet [125] is a deep-learning-based approach that classifies DNA sequences directly from electrical signals. SquiggleNet provides 90% accuracy with real-time processing, which means that SquiggleNet wrongly identifies, on average, 10% of the DNA segments as irrelevant. UNCALLED [126] is another approach that segments raw signals into possible k-mers and uses Ferragina-Manzini (FM) index along with a probabilistic model to search for k-mer matches with the target reference genome. UNCALLED achieves an accuracy of 93.7% with an average detection speed of <50ms per DNA segment. Sigmap [127] has comparable performance to UNCALLED, but it overcomes the applicability limitations of UNCALLED to large genomes by optimizing the raw signal mapping pipeline. This optimization leads to a 4.4× improvement in detection performance. SquiggleFilter [125] matches reads against a target genome using hardware-accelerated dynamic time warping algorithm [119,128]. SquiggleFilter needs fewer signal measurements to identify irrelevant DNA segments compared to UNCALLED, allowing for earlier ejection from the pores. It is computationally cheaper and particularly suited for sequencing in the field, where the computational resources are limited. The computational step of SquiggleFilter can be as fast as 0.027ms per DNA segment, depending on the size of the to-be-compared-with reference genome. Analysis in raw signal space can also provide information that would otherwise be lost during basecalling, such as chemical modifications of nucleotides, or estimating the length of the poly(A) tail of mRNA [122].

### 5.3. PacBio

For PacBio sequencing, 30 hours of continuous fluorescent light pulses are recorded as a movie. The movie can be directly interpreted into bases, resulting in multiple subreads (stored in BAM format), where each subread corresponds to a single pass over the circular DNA fragment. The current PacBio basecalling workflow is CCS [129], which includes aligning the subreads to each other using computationally expensive sequencing alignment tools, such as KSW2 [130] and Edlib [131], and other polishing steps. We observe that there are currently neither hardware acceleration nor alternative more efficient (more customized than KSW2 and Edlib) algorithms for improving the runtime of PacBio basecalling.

## 6. Quality Control

The goal of the quality control (QC) step (**Figure 1.3**) is to examine the quality of some regions in the read sequence or the entire read sequence and trim them if they are of low quality or not needed anymore (as in adapters for sequencing). Some of the causes for low quality regions are library preparation (e.g., fragmenting the DNA into very short fragments) and sequencing (e.g., low base quality during sequencing) [95,132]. The QC step ensures high quality of the reads and hence high quality downstream analysis.

The QC step evaluates the quality of the entire read sequence by (1) examining the intrinsic quality of the read sequence, that is the average quality score generated by the sequencing machine before or after basecalling, (2) assessing the length of the read, and (3) evaluating the number of ambiguous bases (N) in a read. The QC step also evaluates the quality of individual bases for (1) masking out an untrustworthy region in a read if the average quality score of the region's bases is low and (2) trimming beginning or/and trailing regions of a read if it is. For example, Illumina sequences have degraded quality towards the ends of the reads, while adapter sequences are added to both head and tail of the DNA fragment for sequencing. Looking at quality distribution over base positions can help decide the trimming sites. There are also other quality control steps that can be applied during or after read mapping, such as those provided by Picard [133], and can be hardware accelerated as in [134]. Some other quality control steps can be carried out directly after sequencing and before basecalling, such as trimming barcode sequences used for labeling different samples to be sequenced within the same sequencing run [135,136].

There exists a number of software tools for performing QC. FastQC [137] is the most popular tool for controlling the quality of Illumina sequencing data. FastQC takes a FASTQ file as its input and performs ten different quality analyses. A given FASTQ file may pass or fail each analysis and a FASTQ file is usually accepted when it passes all quality analyses. There are also a large number of QC software tools for long read (both nanopore and HiFi), such as LongQC [138], which uses expensive read mapping, minimap2 [130], for low coverage detection. RabbitQC [139] exploits modern multicore CPUs to parallelize the QC computations and provides an order of magnitude faster performance for the three prominent types of sequencing technologies. We observe that there are currently a large number of software tools for QC, but we still lack efficient hardware accelerators for improving the QC step.

## 7. Read Mapping

The goal of read mapping is to locate possible subsequences of the reference genome sequence that are similar to the read sequence while allowing at most *E* edits, where *E* is the *edit distance threshold*. Tolerating a number of differences is essential for correctly finding possible locations of each read due to sequencing errors and genetic variations. Mapping billions of reads to the reference genome is *still* computationally expensive [27,35,36,53,140]. Therefore, most existing read mapping algorithms exploit two key heuristic steps, *indexing* and *filtering*, to reduce the search space for each read sequence in the reference genome. Read mapping includes four computational steps (**Figure 1.4**), indexing, seeding, pre-alignment filtering, and sequence alignment. First, a read mapper starts with building a large index database using subsequences (called *seeds*) extracted from a reference genome to enable quick and efficient querying of the reference genome. Second, the mapper uses the prepared index database to determine one or more possible regions of the reference genome that are likely to be similar to each read sequence by matching subsequences extracted from each read with the subsequences stored in the index database.

Third, the read mapper uses filtering heuristics to quickly examine the similarity for every read sequence and one potential matching segment in the reference genome identified during seeding. As only a few short subsequences are matched between each read sequence and each reference genome segment, there can be a large number of differences between the two sequences. Hence filtering heuristics aim to eliminate most of the dissimilar sequence pairs by performing minimal computations. Fourth, the mapper performs sequence alignment to check whether or not the remaining sequence pairs that pass the filter are actually similar. Due to potential differences, the similarity between a read and a reference sequence segment must be identified using an approximate string matching (ASM) algorithm. The ASM typically uses a computationally-expensive dynamic programming (DP) algorithm to *optimally* (1) examines all possible *prefixes* of two sequences and tracks the prefixes that provide the highest possible *alignment score* (known as *optimal alignment*), (2) identify the type of each difference (i.e., insertion, deletion, or substitution), and (3) locate each difference in one of the two given sequences. Such alignment information is typically output by read mapping into a sequence alignment/map (SAM, and its compressed representation, BAM) file [141]. The alignment score is a quantitative representation of the quality of aligning each base of one sequence to a base from the other sequence. It is calculated as the sum of the scores of all differences and matches along the alignment implied by a user-defined scoring function. DP-based approaches usually have quadratic time and space complexity (i.e., ($m^2$) for a sequence length of *m*), but they avoid re-examining the same prefixes many times by storing the examination results in a DP table. The use of DP-based approaches is unavoidable when optimality of the alignment results is desired [142]. Given the time spent on read mapping, all four steps have been targeted for acceleration.

**Table 3: Evaluation analysis of three state-of-the-art read mappers, BWA-MEM2, minimap2, and pbmm2, for three different types of sequencing data, short reads, ultra-long reads, and accurate long reads, respectively.**

|  | **Short reads** | **Ultra-long reads** | **Accurate long reads** |
|---|---|---|---|
| Read mapping tool | BWA-MEM2 | minimap2 | pbmm2 |
| Version | 2.2.1 | 2.24-r1122 | 1.7.0 |
| Reference genome | Human genome (HG38), GCA_000001405.15, FASTA size 3.2 GB | | |
| Indexing time (CPU) | 2002 sec | 163 sec | 144 sec |
| Indexing peak memory | 72.3 GB | 11.4 GB | 14.4 GB |
| Indexing size* | 17 GB | 7.3 GB | 5.7 GB |
| Read set (accession number) | https://www.ebi.ac.uk/ena/browser/view/ERR194147 | https://ftp-trace.ncbi.nlm.nih.gov/ReferenceSamples/giab/data/AshkenazimTrio/HG003_NA24149_father/UCSC_Ultralong_OxfordNanopore_Promethion/GM24149_1.fastq.gz | https://ftp-trace.ncbi.nlm.nih.gov/ReferenceSamples/giab/data/AshkenazimTrio/HG003_NA24149_father/PacBio_CCS_15kb_20kb_chemistry2/reads/PBmixSequel729_1_A01_PBTH_30hours_19kbV2PD_70pM_HumanHG003.fastq.gz |
| Number of reads | 1,430,362,384 | 6,724,033 | 1,289,591 |
| Number of bases | 144,466,600,784 | 82,196,263,791 | 24,260,611,730 |
| Read mapping time | 868.9 hours[1] | 48.6 hours | 17.5 hours |
| Mapping peak memory | 131.2 GB[1] | 9.6 GB | 20 GB |
| Number of mapped reads[2] | 2,842,576,947 | 3,854,572 | 1,286,256 |
| Mapping throughput (input bases/mapping time) | 46,185 bases/sec | 469,801 bases/sec | 385,090 bases/sec |
| Number of output mappings (SAM) | 2,875,143,231 | 8,322,218 | 1,396,899 |
| File size of output mappings (SAM) | 222.6 GB | 190.2 GB | 54.4 GB |

[1]For paired-end read mapping
[2]After excluding secondary, supplementary, and unmapped alignments using SAMtools and SAM flag 2308.

## 7.1. Accelerating Indexing and Seeding

Indexing and seeding fundamentally use the same algorithm to extract the subsequences from the reference genome or read sequences. The only difference is that the indexing step stores the seeds extracted from the reference genome in an indexing database, while the seeding step uses the extracted seeds to query the indexing database. The indexing step populates a lookup data structure that is indexed by the contents of a seed (e.g., its hash value), and identifies all locations where a seed exists in the reference genome **(Figure 1b)**. Indexing needs to be done only once for a reference genome, thus it is not on the critical path for most bioinformatics applications. Seeding is performed for every read sequence and thus it contributes to the execution time of read mapping. However, the number of extracted seeds in both steps (indexing and seeding), the length of each seed, and the frequency of each seed can significantly impact the overall memory footprint, performance, and accuracy of read mapping [130,143,144]. For example, querying very short seeds leads to a large number of mapping locations that need to be checked for a sequence alignment, which makes later steps more computationally costly. In contrast, querying very long seeds may *prevent* identifying some mapping locations. This is because the querying process usually requires the entire seed to exactly appear in the indexing database, and longer seeds have a higher probability of containing mismatches. This can lead to missing some mapping locations and causing a low accuracy (defined in this context as *sensitivity*, the ability of a read mapper to find the location of a read sequence that already exists in the reference genome). The properties of the data affect the tradeoffs between these choices, for example long reads tend to have a higher error rate, thus shorter seed lengths are appropriate for good sensitivity. There are three major directions for improving the indexing and seeding steps: (1) better seed sampling techniques, (2) better indexing data structures, and (3) accelerating the task and minimizing its data movement through specialized hardware.

### 7.1.1. Sampling Seeds

The goal of sampling the seeds is to reduce redundant information that can be inferred from extracted seeds. For example, choosing all possible overlapping subsequences of length $k$, called $k$-mers, as seeds causes each base to appear in $k$ seeds, causing unnecessarily high memory footprint and inefficient querying due to large number of seed hits. Thus, state-of-the-art read mapping algorithms (e.g., minimap2 [130]) typically aim to reduce the number of seeds that are extracted for the index structure by sampling all possible k-mers into a smaller set of k-mers. A common strategy to choose such a smaller set is to impose an ordering (e.g. lexicographically or by hash value) on every group of $w$ overlapping k-mers and choosing only the k-mer with the smallest order as a seed, known as the minimizer k-mers [145]. Minimizer-based approach guarantees finding at least one seed in a group of k-mers, known as *windowing guarantee*, ensuring low information loss depending on the values of $k$ and $w$. Additionally, the frequency (i.e., the size of the location list) of each minimizer seed can be restricted up to a certain threshold to reduce the workload for querying and filtering the seed hits [143,146,147]. Similar to minimizers, the syncmer approach [148] is a more recent type of sampling method with a different selection criteria than minimizers that has shown to provide more uniform distribution of seeds to achieve better sensitivity. The syncmer approach chooses a seed as a minimizer whose substring located at a fixed location achieves the smallest order compared to that of substrings of other seeds. This strategy ensures a certain gap between any two consecutive minimizers, which enables read mappers to report mapping locations for reads that are unmapped using minimizers-based read mappers [149].

To increase the sensitivity of read mappers, other approaches can be applied, such as spaced [150] and strobemer [151] seeds. Spaced seeds [150] exclude some bases from each seed following a predetermined pattern, where the resulting seeds are composed of multiple shorter substrings [152]. Spaced seeds can achieve high sensitivity when finding seed matches by allowing the excluded bases to mismatch (substitute) with their corresponding bases. S-conLSH [153] is a recent approach that applies the locality-sensitive hashing idea and uses multiple patterns on the same sequence to enable excluding different sets of characters belonging to a sequence. A recent approach, known as strobemers [151], improves on spaced seeds by varying the sizes of the spaces dynamically based on selection criterias. These strategies for joining/linking seeds are orthogonal to minimizers and syncmers, and recent work shows such approaches can be combined for additional sensitivity improvements [154].

There are also other works that use the minimizer sampling strategy without providing a windowing guarantee, such as MinHash [155]. MinHash finds a single minimizer k-mer from an entire sequence (i.e., *w* equals the sequence length). To find many minimizer k-mers from the same sequence, the idea is to use many hash functions and find the minimizer k-mer from each hash function. The goal is to find a minimizer k-mer at *n*-many regions of a sequence using *n*-many different hash functions, providing a sampled set of k-mers. Although the MinHash approach is effective when comparing sequences of similar lengths, it uses many redundant minimizers when comparing sequences of varying length to ensure high accuracy, which comes with a high cost of performance and memory overhead [156].

**7.1.2. Improving Data Structures for Seed Lookups**
After choosing the appropriate method for extracting seeds, the goal is to store or query them using the index. The straightforward data structure to find seed matches is a hash table that stores the hash value of each seed as a key, and location lists of each seed as values. Hash tables have been used since 1988 in read mapping [18] and are still used even by the state-of-the-art read mappers as they show good performance in practice [130]. Choosing a hash function is an important design choice for hash tables. It is desired to use hash functions with low collision rates so that different seeds are not assigned to the same hash value. It is also desired to increase the collision rate for highly similar seeds to improve the overall sensitivity. A recent approach, BLEND [144], aims to generate hash values such that highly similar seeds can have the same hash value while dissimilar seeds are still assigned to different hash values with low collision rates. Such an approach can find approximate (i.e., fuzzy) matches of seeds directly using their hash values, which can be applied to other seeding approaches that enable finding inexact matching, such as spaced seeds and strobemers.

There are also other data structures that can be efficiently used with the aim of reduced memory footprint and improved querying time. One example of such data structures is FM-index (implemented by Langarita et al. [157]), which provides a compressed representation of the full-text index, while allowing for querying the index without the need for decompression. This approach has two main advantages. 1) We can query seeds of arbitrary lengths, which helps to reduce the number of queried seeds. 2) It typically has less (by 3.8×) memory footprint compared to that of the indexing step of minimap2 [18]. However, there is no significant difference in read mapping runtime due to the use of either indexing data structure [18]. One major bottleneck of FM-indexes is that locating the exact matches by querying the FM-index is significantly slower than that of classical indexes [157,158]. The FM-index can be accelerated by at least 2× using SIMD-capable CPUs [159]. BWA-MEM2 [158] proposes an uncompressed version of the FM-index that is at least 10× larger than the compressed FM-index to speed up the querying step by 2×. The seeding step of BWA-MEM2 can be further accelerated by 2× by using *enumerated radix trees* on recent CPUs to reduce the number of memory accesses and improve the access patterns [160]. Hash-based minimizer lookup can be replaced with learned indexes [161]. The learned indexes use machine learning models to predict the locations of the queried minimizers. The expected benefit of such a machine learning-based index is the reduced size of the index as it does not store the locations of the seeds. However, it is shown that a learned-index based read mapper has the same memory footprint as the hash-table based read mapper [162].

### 7.1.3. Reducing Data Movement During Indexing

Indexing and seeding remain memory-intensive tasks [53], and hence do not fit modern processor centric systems well. An alternative approach is PIM, where processing happens either inside the memory chip itself, or close to it [34]. This approach can improve both energy efficiency, by moving data a shorter distance, as well as throughput, by providing more total memory bandwidth [34]. MEDAL [163] proposes integrating small ASIC accelerators for seeding close to off-the-shelf DRAM chips on standard LRDIMM memory modules. GenStore [53] proposes a seeding and filtering accelerator inside SSDs, providing comparable advantages to PIM approaches, but with the key difference that the index and reads do not have to be moved outside of the storage device for seeding. The higher internal bandwidth of SSDs provides increased throughput, and the reduced data movement improves energy efficiency. RADAR [164] implements the search for exact matches in an index database by storing the database in 3D Resistive Random Access Memory (ReRAM) based Content Addressable memory (ReCAM). The database can be directly queried without offloading it, leading to a small amount of data movement and highly energy efficient operation.

### 7.2. Accelerating Pre-Alignment Filtering

After finding one or more potential mapping locations of the read in the reference genome, the read mapper checks the similarity between each read and each segment extracted at these mapping locations in the reference genome. These segments can be *similar* or *dissimilar* to the read, though they share common seeds. To avoid examining dissimilar sequences using computationally-expensive sequence alignment algorithms, read mappers typically use filtering heuristics that are called *pre-alignment filters*. The key idea of pre-alignment filtering is to quickly estimate the number of edits between two given sequences and use this estimation to decide whether or not the computationally-expensive DP-based alignment calculation is needed — if not, a significant amount of time is saved by avoiding DP-based alignment. If two genomic sequences differ by more than the edit distance threshold, then the two sequences are identified as dissimilar sequences and hence DP calculation is not needed. Edit distance is defined as the minimum number of single character changes needed to convert a sequence into the other sequence [165]. In practice, only genomic sequence pairs with an edit distance less than or equal to a user-defined threshold (i.e., $E$) provide useful data for most genomic studies [18,51,52,140,166,167]. Pre-alignment filters use one of four major approaches to quickly filter out the dissimilar sequence pairs: (1) the pigeonhole principle, (2) base counting, (3) $q$-gram filtering, or (4) sparse DP. Long read mappers typically use $q$-gram filtering or sparse DP, as their performance scales linearly with read length and independently of the edit distance.

### 7.2.1. Pigeonhole Principle

The pigeonhole principle states that if $E$ items are put into $E$+1 boxes, then one or more boxes would be empty. This principle can be applied to detect dissimilar sequences and discard them from the candidate sequence pairs used for ASM. If two sequences differ by $E$ edits, then they should share at least a single subsequence (free of edits) among any set of $E$+1 non-overlapping subsequences [140], where $E$ is the edit distance threshold. Pigeonhole-based pre-alignment filtering can accelerate read mappers even without specialized hardware. For example, the Adjacency Filter [147] accelerates sequence alignment by up to 19×. The accuracy and speed of pre-alignment filtering with the pigeonhole principle have been rapidly improved over the last seven years. Shifted Hamming Distance (SHD) [167] uses SIMD-capable CPUs to provide high filtering speed, but supports a sequence length up to only 128 base pairs due to the SIMD register widths. GateKeeper [166] utilizes the large amounts of parallelism offered by FPGA architectures to accelerate SHD and overcome such sequence length limitations. MAGNET [168] provides a comprehensive analysis of all sources of filtering inaccuracy of GateKeeper and SHD. Shouji [140] leverages this analysis to improve the accuracy of pre-alignment filtering by up to two orders of magnitude compared to both GateKeeper and SHD, using a new algorithm and a new FPGA architecture.

SneakySnake [51] achieves up to four orders of magnitude higher filtering accuracy compared to GateKeeper and SHD by mapping the pre-alignment filtering problem to the single net routing (SNR) problem in VLSI chip layout. SNR finds the shortest routing path that interconnects two terminals on the boundaries of a VLSI chip layout in the presence of obstacles. SneakySnake is the only pre-alignment filter that efficiently works on CPUs, GPUs, and FPGAs. To further reduce data movements, SneakySnake is redesigned to exploit the near-memory computation capability on modern FPGA boards equipped with high-bandwidth memory (HBM) [54]. Near-memory pre-alignment filtering improves performance and energy efficiency by 27.4× and 133×, respectively, over SneakySnake running on a 16-core (64 hardware threads) IBM POWER9 CPU [54]. GenCache [169] proposes to perform highly-parallel pre-alignment filtering inside the CPU cache to reduce data movement and improve energy efficiency, with about 20% cache area overhead. GenCache shows that using different existing pre-alignment filters together (a similar approach to [170]), each of which operates only for a given edit distance threshold (e.g., using SHD only when is between 1 and 5), provides a 2.5× speedup over GenCache with a single pre-alignment filter. Several pigeonhole principle based pre-alignment filters are evaluated for wide-range FPGA platforms [171].

### 7.2.2. Base Counting
The base counting filter compares the numbers of bases (A, C, G, T) in the read with the corresponding base counts in the reference segment. The sum of absolute differences of the base counts provides an upper bound on the edit distance of the read and reference segment. If one sequence has, for example, three more Ts than another sequence, then their alignment has at most three edits. If half of the sum of absolute differences between the four base counts is greater than $E$, then the two sequences are dissimilar and the reference segment is discarded. The base counting filter is used in mrsFAST-Ultra [172] and GASSST [170]. Such a simple filtering approach rejects a significant fraction of dissimilar sequences (e.g., 49.8%–80.4% of sequences, as shown in GASSST [170]) and thus avoids a large fraction of expensive verification computations required by sequence alignment algorithms. A PIM implementation of base counting can improve filtering time by 100× compared to its CPU implementation [173].

### 7.2.3. *q*-gram Filtering Approach
The *q*-gram filtering approach considers all of the sequence's possible overlapping substrings of length *q* (known as *q*-grams). Given a sequence of length *m*, there are $m-q+1$ overlapping *q*-grams that are obtained by sliding a window of length *q* over the sequence. A single difference in one of the sequences can affect at most *q* overlapping *q*-grams. Thus, differences can affect no more than $q \cdot E$ *q*-grams, where $E$ is the edit distance threshold. The *minimum number* of shared *q*-grams between two similar sequences is therefore $(m-q+1)-(q \cdot E)$. This filtering approach requires very simple operations (e.g., sums and comparisons), which makes it attractive for hardware acceleration, such as in GRIM-Filter [52]. GRIM-Filter exploits the high memory bandwidth and computation capability in the logic layer of 3D-stacked memory to accelerate *q*-gram filtering in the DRAM chip itself, using a new representation of reference genome that is friendly to in-memory processing. *q*-gram filtering is generally robust in handling only a small number of edits, as the presence of edits in any *q*-gram is significantly underestimated (e.g., counted as a single edit) [174]. The data reuse in GRIM-Filter can be exploited for improving both performance and energy efficiency of filtering [175].

### 7.2.4. Sparse Dynamic Programming
Sparse DP algorithms exploit the exact matches (seeds) shared between a read and a reference segment to reduce execution time. These algorithms exclude the corresponding locations of these seeds from estimating the number of edits between the two sequences, as they were already detected as exact matches during indexing. Sparse DP filtering techniques link the overlapping seeds together to build longer chains and use the total length of the calculated chains as a metric for filtering the sequence pairs. This approach is also known as *chaining*, and is used in minimap2 [130] and rHAT [176]. GPU and FPGA accelerators [177] can achieve 7× and 28× acceleration, respectively, compared to the sequential implementation (executed with 14 CPU threads) of the chaining algorithm used in minimap2. mm2-fast [162] accelerates minimap2's chaining step by up to 3.1× with SIMD instructions. mm2-ax [178] accelerates mm2-fast's chaining step by up to 12.6× using a GPU. Modular Aligner [179] introduces an alternative to chaining based on two seed filtering techniques, achieving better performance than minimap2 in terms of both accuracy and runtime.

### 7.3. Accelerating Sequence Alignment

After filtering out most of the mapping locations that lead to dissimilar sequence pairs, read mapping calculates the sequence alignment information for every read and reference segment extracted at each mapping location. Sequence alignment calculation is typically accelerated using one of two approaches: (1) accelerating optimal affine gap scoring DP-based algorithms using hardware accelerators, and (2) developing heuristics that sacrifice the optimality of the alignment score solution in order to reduce alignment time. Affine gap scores are typically calculated using the Smith-Waterman-Gotoh algorithm [180], allowing for linear integer scores for matches/substitutions, and affine integer scores for gaps. Affine gap scores are more general than linear or unit (edit distance) costs, but are more costly to compute by a constant factor. Despite more than three decades of attempts to accelerate sequence alignment, the fastest known edit distance algorithm [181] has a nearly quadratic running time, $O(m^2/\log_2 m)$ for a sequence of length $m$, which is proven to be a tight bound, assuming the strong exponential time hypothesis holds [142]. A common approach to reducing the algorithmic work without sacrificing optimality is to define an *edit distance threshold*, limiting the maximum number of allowed single-character edits in the alignment. In this case, only a subset of the entries of the DP table is computed, called *diagonal vectors*, as first proposed in Ukkonen's banded algorithm [182]. The number of diagonal vectors required for computing the DP matrix is $2E+1$, where $E$ is the edit distance threshold. This reduces the runtime complexity to $O(m*E)$. This approach is effective for short reads, where the typical sequencing error rates are low, thus a low edit distance threshold can be chosen. Unfortunately, as long reads have high sequencing error rates (up to 20% of the read length), the edit distance threshold for long reads has to be high, which results in calculating more entries in the DP matrix compared to that of short reads. The use of heuristics (i.e., the second approach) helps to reduce the number of calculated entries in the DP matrix and hence allows both the execution time and memory footprint to grow only linearly or less with read length (as opposed to quadratically with classical DP). Next, we describe the two approaches in detail.

### 7.3.1. Accurate Alignment Accelerators

From a hardware perspective, sequence alignment acceleration has five directions: (1) using SIMD-capable CPUs, (2) using multicore CPUs and GPUs, (3) using FPGAs, (4) using ASICs, and (5) using processing-in-memory architectures. Parasail [183] and KSW2 (used in minimap2 [130]) exploit both Ukkonen's banded algorithm and SIMD-capable CPUs to compute banded alignment for a sequence pair with a configurable scoring function. SIMD instructions offer significant parallelism to the matrix computation by executing the same vector operation on multiple operands at once. mm2-fast [162] accelerates KSW2 by up to 2.2× by matching its SIMD capability to recent CPU architectures. KSW2 is nearly as fast as Parasail when KSW2 does not use heuristics (explained in **Section 7.3.2**). The wavefront algorithm (WFA) [184] reformulates the classic Smith-Waterman-Gotoh recursion such that the runtime is reduced to O($m*s$) for a sequence pair of length $m$ and affine gap cost of $s$ without fixing the vale of $s$ ahead of time. It is SIMD-friendly and shows significant speedups for sequence pairs that have high similarity. The memory footprint and runtime complexity of WFA can be improved to O($s^2$) [185]. However, this improved runtime is not practical due to a large constant factor. The memory footprint of the WFA algorithm can be improved to O($s$) at the expense of an increase in runtime complexity to O($m*s$) [186]. LEAP [187] formulates what can be considered a more general version of WFA, which is applicable to any convex penalty scores.

The multicore architecture of CPUs and GPUs provides the ability to compute alignments of many independent sequence pairs concurrently. GASAL2 [188] exploits the multicore architecture of both CPUs and GPUs for highly-parallel computation of sequence alignment with a user-defined scoring function. Unlike other GPU-accelerated tools, GASAL2 transfers the bases to the GPU, without encoding them into binary format, and hides the data transfer time by overlapping GPU and CPU execution. GASAL2 is up to 20× faster than Parasail (when executed with 56 CPU threads). BWA-MEM2 [158] accelerates the banded sequence alignment of its predecessor (BWA-MEM [189]) by up to 11.6×, by leveraging multicore and SIMD parallelism. A GPU implementation [190] of the WFA algorithm improves the original CPU implementation by 1.5-7.7× using long reads.

Other designs, such as FPGASW [191], exploit the very large number of hardware execution units in FPGAs to form a linear systolic array [192]. Each execution unit in the systolic array is responsible for computing the value of a single entry of the DP matrix. The systolic array computes a single vector of the matrix at a time. The data dependency between the entries restricts the systolic array to computing the vectors sequentially (e.g., top-to-bottom, left-to-right, or in an anti-diagonal manner). FPGASW has a similar execution time as its GPU implementation, but is 4× more power efficient. SeedEx [193] designs an FPGA accelerator similar to FPGASW, but improves hardware utilization by speculatively computing fewer than 2E+1 diagonal bands, and then applying optimality checking heuristics to guarantee correct results. An FPGA accelerator [194] can accelerate the WFA algorithm by up to 8.8× and improve its energy efficiency by 9.7× for only short reads.

Specialized hardware accelerators (i.e., ASIC designs) provide application-specific, power- and area-efficient solutions to accelerate sequence alignment. For example, GenAx [195] is composed of SillaX, a sequence alignment accelerator, and a second accelerator for finding seeds. SillaX supports both a configurable scoring function and traceback operations. SillaX is more efficient for short reads than for long reads, as it consists of an automata processor whose performance scales quadratically with the edit distance. GenAx is 31.7× faster than the predecessor of BWA-MEM2 (i.e., BWA-MEM [189]) for short reads. Recent PIM architectures such as RAPID [196] exploit the ability to perform computation inside or near the memory chip to enable efficient sequence alignment. RAPID modifies the DP-based alignment algorithm to make it friendly to in-memory parallel computation by calculating two DP matrices (similar to Smith-Waterman-Gotoh [180]): one for calculating substitutions and exact matches and another for calculating insertions and deletions. RAPID claims that this approach efficiently enables higher levels of parallelism compared to traditional DP algorithms. The main two benefits of RAPID and such PIM-based architectures are higher performance and higher energy efficiency [33,34], as they alleviate the need to transfer data between the main memory and the CPU cores through slow and energy hungry buses, while providing high degree of parallelism with the help of PIM. RAPID is on average 11.8× faster and 212.7× more power efficient than 384-GPU cluster implementation of sequence alignment, known as CUDAlign [197]. A recent PIM architecture of WFA algorithm implemented in real hardware provides up to 4.87× higher throughput than the 56-thread CPU implementation using short reads [198].

### 7.3.2. Alignment Accelerators with Limited Functionality

The second direction is to *limit* the functionality of the alignment algorithm or *sacrifice* the optimality of the alignment solution in order to reduce execution time. The use of restrictive functionality and heuristics limits the possible applications of the algorithms that utilize this direction. Examples of limiting functionality include limiting the scoring function (e.g. allowing only linear gap or unit scores), and calculating only the alignment score without performing the backtracking step [199]. There are several existing algorithms and corresponding hardware accelerators that limit scoring function flexibility. An example of limiting the scoring function is Myers' bit-vector algorithm [200], where the scoring function is limited to edit distance [165]. In this case, all types of edits are penalized equally when calculating the total alignment score. Restrictive scoring functions enable computation with smaller bit-widths, such that either smaller registers can be used, or multiple DP entries fit into a single SIMD register. This reduces the total execution time of the alignment algorithm by operating on multiple DP entries in parallel in a SIMD fashion.In the case of Myer's bit-vector algorithm a single 64-bit register can hold the values of 64 entries of the DP matrix. BitPAl [201] expands on the idea by limiting the scoring function to linear gap scores and achieves speedups through bit-parallel execution. ASAP [202] accelerates edit distance calculation by up to 63.3× using FPGAs compared to its CPU implementation. The use of a fixed scoring function as in Edlib [131], which is the state-of-the-art implementation of Myers' bit-vector algorithm, helps to outperform Parasail (which uses a flexible scoring function) by 12–1000×. One downside of a limited scoring function is that it may lead to the selection of a suboptimal sequence alignment, relative to an affine gap scoring function as in the Smith-Waterman-Gotoh algorithm. There are also a large number of edit distance approximation algorithms that provide a reduction in time complexity (e.g., $m^{1.647}$ instead of $m^2$), but they suffer from providing overestimated edit distance [203–206].

There are other algorithms and hardware architectures that provide low alignment time by trading off accuracy. Darwin [36] builds a customized hardware architecture to speed up the alignment process, by dividing the DP matrix into overlapping submatrices and greedily processing each submatrix using systolic arrays. Darwin provides three orders of magnitude speedup compared to Edlib [131]. Greedily processing each submatrix reduces the number of calculated entries of the full DP matrix and hence reduces the memory footprint and algorithmic workload, but it leads to suboptimal alignment calculation [170]. Darwin claims that choosing a large submatrix size (≥ 320 × 320) and ensuring sufficient overlap (≥128 entries) between adjacent submatrices may provide optimal alignment calculation for some datasets. GenASM [35] is a framework that uses bit-vector-based ASM and a similar strategy as Darwin to accelerate multiple steps of the genome analysis pipeline, and is designed to be implemented inside 3D-stacked memory. Through a combination of hardware–software co-design to unlock parallelism, and processing-in-memory to reduce data movement, GenASM achieves 111×/116× speedup over state-of-the-art software read mappers while reducing power consumption by 33×/37×.

There are other proposals that limit the number of calculated entries of the DP matrix based on one of two approaches: (1) using sparse DP or (2) using a greedy approach to maintain a high alignment score. Both approaches suffer from producing possibly suboptimal alignments [207,208]. The first approach uses the same sparse DP algorithm used for pre-alignment filtering but as an alignment step, as done in the exonerate tool [207]. This includes applying DP-based alignment algorithms only between every two non-overlapping chains to quickly estimate the total number of edits. The second approach is employed in *X*-drop [208], which (1) avoids calculating entries (and their neighbors) whose alignment scores are more than below the highest score seen so far (where is a user-specified parameter), and (2) stops early when a high alignment score is not possible. The *X*-drop algorithm is guaranteed to find the optimal alignment between relatively-similar sequences for *only some* scoring functions [208]. A similar algorithm (known as *Z*-drop) makes KSW2 at least 2.6× faster than Parasail. A recent GPU implementation [209] of the *X*-drop algorithm is 3.1–120.4× faster than KSW2. A related approach is *adaptive banding* [210] (and its improved algorithm Block aligner [211]), where the band of calculated diagonals is shifted up or down depending on the highest score in the last calculated anti-diagonal.

## 8. Variant Calling

The goal of variant calling (**Figure 1.5**) is to find the differences between an individual or a group of individuals (i.e., population) compared to a reference genome of a species. Calling the variants is an essential step in genome analysis because the attributes of an individual or a population (e.g., phenotypes or diseases) are determined from these variations. To determine these variants, there are several steps that need to be performed as outlined by tools such as GATK's best practices [212] and DeepVariant's workflow [213]. Variant calling usually iterates over the read mapping information to identify the variants such as SNPs, short INDELs and SVs. Although there are several algorithms to find SNPs and short INDELs, the main idea of most of these tools is to find the locations in a genome where the reference genome and the sequencing reads have different bases. To find such regions, several mapped reads should consistently show the same edit at the same position of a reference genome to call the variant with a high quality.

Calling the variants with high quality is essential for performing accurate downstream analysis (e.g., validating a genetic disease) [214]. There are several parameters that contribute to calling high quality variants such as high sequencing depth of coverage, highly accurate sequencing reads (i.e., Illumina and PacBio HiFi), long reads, accurate and deterministic read mappers. The sequencing depth of coverage refers to the average number of reads mapped to each location in the reference genome. This helps variant callers to better distinguish the genetic mutations from errors (both sequencing errors and read mapping artifacts [31]). It is also known that variant calling tools may also be nondeterministic such that running the same tool multiple times may result in different results [31]. Thus, it is essential for the community to provide the best practices to achieve high accuracy due to many parameters involved in high quality variant calling.

There are several efforts in the field to provide best practices when performing genomic analysis that includes variant calling. One of the efforts is to provide benchmarking studies to evaluate the accuracy of the variant calling output [215]. Such comparisons are usually done by comparing the output from a variant caller with a ground truth dataset (e.g., GiAB [216]). Another effort is to suggest the best pipeline of tools to achieve the best accuracy for variant calling [214]. Another effort focuses on identifying the computational bottlenecks in the best practices for variant calling and accelerates these bottlenecks to achieve high performance in variant calling [217,218]. Compared to the number of proposed software and hardware accelerators for read mapping, there are only a limited number of hardware accelerators for variant calling. Given the large execution time (**Table 4**) of this key step in genome analysis, there is still a huge need for accelerating state-of-the-art variant calling tools, such as DeepVariant.

DeepVariant has 3 key steps: 1) processing mapping data (called *making examples*), 2) variant classification, and 3) generating variant calls. The first step prepares read mapping output data (e.g., read bases, base quality, edit information, strand information) for the neural network. The output of read mapping (i.e., SAM file) has also some irrelevant information that needs to be cleaned. This includes identifying and removing read duplicates that can be a result of library preparation using PCR, as they do not lead to any useful information for variant calling. The second step is when the deep neural network does the classification. The last step interprets the classification output by the neural network as variant calls stored in *VCF* format. Information about each step can be found in [219]. Based on Table 4, we make three key observations. 1) The first step of DeepVariant consumes about 50% of the execution time of the second step, variant classification. The first stage can be accelerated by 2× using modern FPGAs as in [134]. 2) Using nearly 10× more bases of short read mapping data compared to that of accurate long reads leads to nearly the same number of variant calls (PASS) and half of the number of called variants (RefCALL). 3) Variant calling using ultra-long reads is computationally very expensive, which can be mainly because of the errors in the raw sequencing data, as we discuss in **Section 3.4**.

**Table 4: Evaluation analysis of variant calling, using DeepVariant tool, using read mapping results of three different types of sequencing data, short reads, ultra-long reads, and accurate long reads.**

|  | **Short reads** | **Ultra-long reads** | **Accurate long reads** |
|---|---|---|---|
| Variant calling tool | DeepVariant | PEPPER + DeepVariant | DeepVariant |
| Version | 1.3.0 | 0.8 | 1.3.0 |
| Total number of bases in input SAM file | 250,103,434,512 | 56,958,985,752 | 23,944,354,059 |
| Phase 1 (making examples) CPU Time (sec) | 250,359 | 1,136,356 | 230,066 |
| Phase 2 (calling variants) CPU Time (sec) | 473,962 | 3,480,419 | 549,272 |
| Phase 3 (post-processing variants) CPU Time (sec) | 2,193 | 2,765 | 6,201 |
| Total Run Time (sec) | 746,514 | 4,619,540 | 785,539 |
| Number of Called Variants (PASS) | 4,644,980 | 6,054,168 | 4,589,024 |
| Number of Called Variants (RefCall) | 1,313,292 | 6,722,265 | 2,603,968 |
| Variant Calling Throughput (Number of called variants / sec) | 7.9 variants/sec | 2.77 variants/sec | 9.2 variants/sec |

## 9. Discussion and Future Opportunities

Despite more than three decades of attempts, bridging the performance gap between sequencing machines and computational analysis is still challenging. We summarize six main challenges below.

First, we need to accelerate the entire genome analysis process rather than its individual steps. Accelerating only a single step of genome analysis limits the overall achieved speedup according to Amdahl's Law. However, some of the computational steps included in genome analysis pipeline are also included in other genomics pipelines. For example, improving read mapping performance positively impacts almost all genomic analyses that use sequencing data [13,27,35,36,53]. The contribution of read mapping to the entire analysis pipeline varies depending on the application. For example, read mapping takes up to 1) 45% of the execution time when discovering sequence variants in cancer genomics studies [220], and 2) 60% of the execution time when profiling the taxonomy of a multi-species (i.e., metagenomic) sample [13]. Illumina and NVIDIA started following a more holistic approach, and they claim to accelerate genome analysis by more than 48×, mainly by using specialization and hardware/software co-design. Illumina has built an FPGA-based platform, called DRAGEN [221], that accelerates all steps of genome analysis, including read mapping and variant calling. DRAGEN reduces the overall analysis time from 32 CPU hours to only 37 minutes [222]. NVIDIA has built Parabricks, a software suite accelerated using the company's latest GPUs. Parabricks [223] can analyze whole human genomes at 30× coverage in about 45 minutes.

Second, we need to reduce the high amount of data movement that takes place during genome analysis. Moving data (1) between compute units and main memory, (2) between multiple hardware accelerators, and (3) between the sequencing machine and the computer performing the analysis incurs high costs in terms of execution time and energy. These costs are a significant barrier to enabling efficient analysis that can keep up with sequencing technologies, and some recent works try to tackle this problem [33,34,52]. DRAGEN reduces data movement between the sequencing machine and the computer performing analysis by adding specialized hardware support inside the sequencing machine for data compression. However, this still requires movement of compressed data. GenStore [53] mitigates data movement from the storage devices to the rest of the system (processors and main memory) by processing more than 80% of the input read set inside the storage device.

Third, we need to build more specialized hardware accelerators that are mainly developed for genomics. Computer programs are currently widely used for analyzing genomic data, which limits their scaling capability and efficiency to handle population-level analyses. We are witnessing a paradigm shift to near-data computing with more specialized hardware accelerators for other key applications such as self-driving cars [224], and artificial intelligence with the largest chip ever [225]. This already fuelled huge interest in genomics especially from large companies, such as NVIDIA, which introduces GPU H100 boards that are equipped with HBM3 and hardware support for building and calculating DP matrix for sequence alignment. UPMEM also shows significant benefits for using their PIM-capable memory devices for genome analysis [226]. We envision that performing genome analysis inside the sequencing machine itself using emerging technologies (e.g., PIM-enabled systems) can significantly improve efficiency by eliminating sequencer-to-computer movement, and embedding a single specialized chip for genome analysis within a portable sequencing device can potentially be a key enabler for new applications of genome sequencing (e.g., rapid surveillance of diseases such as Ebola [7] and COVID-19 [6,227], near-patient testing, bringing precision medicine to remote locations). Unfortunately, efforts in this direction remain very limited.

Fourth, an emerging problem with using a single reference genome for an entire species is the reference genome bias. The use of a single reference genome can bias the mapping process and downstream analysis towards the DNA composition and variations present in the reference organism due to population-specific genetic variations, individual's genetic variations, and sequencing errors [228,229]. An emerging technique to overcome reference bias is the use of graph-based representations of a species' genome, known as genome graphs [230]. A genome graph represents the reference genome and known genetic variations in the population as a graph-based data structure. Genome graphs are growing in popularity for genome analysis, which requires modifying existing tools or introducing new tools and hardware accelerators for supporting genome graphs instead of linear representations of reference genomes. Hardware acceleration is demonstrated to greatly benefit sequence mapping to genome graphs. SeGraM is the *first* hardware acceleration framework for sequence-to-graph mapping and alignment, where it provides an order of magnitude faster and more energy efficient performance compared to software sequence-to-graph mapping tools [231]. A new direction to alleviate the computation overhead of using different reference genomes is to update the existing results of one step of the genome analysis pipeline for the new reference genome without re-running the step's algorithm again. The efforts in this direction are still limited to only read mapping [61,62,232].

Fifth, we need to develop flexible hardware architectures that do not conservatively limit the range of supported parameter values at design time. Commonly-used read mappers (e.g., minimap2) have different input parameters, each of which has a wide range of input values. For example, the edit distance threshold is typically user defined and can be very high (15-20% of the read length) for recent long reads. A configurable scoring function is another example, as it determines the number of bits needed to store each entry of the DP matrix (e.g., DRAGEN imposes a restriction on the maximum frequency of seed occurrence). Due to rapid changes in sequencing technologies (e.g., high sequencing error rate and longer read lengths) [96,98], these design restrictions can quickly make specialized hardware obsolete. Thus, read mappers need to adapt their algorithms and their hardware architectures to be modular and scalable so that they can be implemented for any sequence length and edit distance threshold based on the sequencing technology.

Sixth, we need to adapt existing genomic data formats for hardware accelerators or develop more efficient file formats. Most sequencing data is stored in the FASTQ/FASTA format, where each base takes a single byte (8 bits) of memory. This encoding is inefficient, as only 2 bits (3 bits when the ambiguous base, N, is included) are needed to encode each DNA base. The sequencing machine converts sequenced bases into FASTQ/FASTA format, and hardware accelerators convert the file contents into unique (for each accelerator) compact binary representations for efficient processing. This process that requires multiple format conversions wastes time. For example, only 43% of the sequence alignment time in BWA-MEM2 [158] is spent on calculating the DP matrix, while 33% of the sequence alignment time is spent on pre-processing the input sequences for loading into SIMD registers, as provided in [158]. To address this inefficiency, we need to widely adopt efficient hardware-friendly formats, such as UCSC's 2bit format (https://genome.ucsc.edu/goldenPath/help/twoBit), to maximize the benefits of hardware accelerators and reduce resource utilization. We are not aware of any recent read mapper that uses such formats.

Looking into the late future, even if accurately sequencing the entire genome as a single string might be possible, we believe that most of the tools and hardware accelerators involved in the intelligent genome analysis pipeline will continue to remain a crucial component in analyzing and comparing the sequencing data. For example, we still need to quickly and efficiently compare complete genomes together for inferring variations and identifying metagenomic taxonomy profiles. The acceleration efforts we highlight in this work represent state-of-the-art efforts to reduce current bottlenecks in the genome analysis pipeline. We hope that these efforts and the challenges we discuss provide a foundation for future work in making genome analysis faster, more accurate, privacy-preserving, more energy-efficient, and cost-effective; *simply more intelligent*.

## Acknowledgments

We thank the SAFARI group members for feedback and the stimulating intellectual environment. We acknowledge the generous gifts and support provided by our industrial partners: Google, Huawei, Intel, Microsoft, VMware, and the Semiconductor Research Corporation. M.A. dedicates this paper to the memory of his father, who passed away on 9th March 2022.

[225] G. Lauterbach, The Path to Successful Wafer-Scale Integration: The Cerebras Story, IEEE Micro. 41 (2021) 52–57.

[226] D. Lavenier, R. Cimadomo, R. Jodin, Variant Calling Parallelization on Processor-in-Memory Architecture, in: 2020 IEEE International Conference on Bioinformatics and Biomedicine (BIBM), 2020: pp. 204–207.

[227] M. Alser, J.S. Kim, N.A. Alserr, S.W. Tell, O. Mutlu, COVIDHunter: An Accurate, Flexible, and Environment-Aware Open-Source COVID-19 Outbreak Simulation Model, arXiv [q-bio.PE]. (2021). http://arxiv.org/abs/2102.03667.

[228] R.M. Sherman, J. Forman, V. Antonescu, D. Puiu, M. Daya, N. Rafaels, M.P. Boorgula, S. Chavan, C. Vergara, V.E. Ortega, A.M. Levin, C. Eng, M. Yazdanbakhsh, J.G. Wilson, J. Marrugo, L.A. Lange, L.K. Williams, H. Watson, L.B. Ware, C.O. Olopade, O. Olopade, R.R. Oliveira, C. Ober, D.L. Nicolae, D.A. Meyers, A. Mayorga, J. Knight-Madden, T. Hartert, N.N. Hansel, M.G. Foreman, J.G. Ford, M.U. Faruque, G.M. Dunston, L. Caraballo, E.G. Burchard, E.R. Bleecker, M.I. Araujo, E.F. Herrera-Paz, M. Campbell, C. Foster, M.A. Taub, T.H. Beaty, I. Ruczinski, R.A. Mathias, K.C. Barnes, S.L. Salzberg, Assembly of a pan-genome from deep sequencing of 910 humans of African descent, Nat. Genet. 51 (2019) 30–35.

[229] S. Ballouz, A. Dobin, J.A. Gillis, Is it time to change the reference genome?, Genome Biol. 20 (2019) 159.

[230] B. Paten, A.M. Novak, J.M. Eizenga, E. Garrison, Genome graphs and the evolution of genome inference, Genome Res. 27 (2017) 665–676.

[231] D.S. Cali, K. Kanellopoulos, J. Lindegger, Z. Bingöl, G.S. Kalsi, Z. Zuo, C. Firtina, M.B. Cavlak, J. Kim, N.M. Ghiasi, G. Singh, J. Gómez-Luna, N.A. Alserr, M. Alser, S. Subramoney, C. Alkan, S. Ghose, O. Mutlu, SeGraM: A Universal Hardware Accelerator for Genomic Sequence-to-Graph and Sequence-to-Sequence Mapping, arXiv [cs.AR]. (2022). http://arxiv.org/abs/2205.05883.

[232] J.S. Kim, C. Firtina, M.B. Cavlak, D.S. Cali, C. Alkan, O. Mutlu, FastRemap: A Tool for Quickly Remapping Reads between Genome Assemblies, arXiv [q-bio.GN]. (2022). http://arxiv.org/abs/2201.06255.